\documentclass[aps,prb,twocolumn,showpacs,twoside,amsmath, amsfonts]{revtex4}
\usepackage{psfrag}
\usepackage{graphicx}
\usepackage{epsfig}
\usepackage{amssymb}

\begin{document}
\title{
 Exact solution for eigenfunction statistics at the center-of-band anomaly in the Anderson localization model}
\author{V.E.Kravtsov$^{1,2}$ and V.I.Yudson$^{3}$}
\affiliation{$^{1}$The Abdus Salam International Centre for
Theoretical Physics, P.O.B. 586, 34100 Trieste, Italy.\\
$^{2}$Landau Institute for Theoretical Physics, 2 Kosygina
st.,117940 Moscow, Russia.\\$^{3}$Institute for Spectroscopy,
Russian Academy of Sciences, 142190 Troitsk, Moscow reg., Russia.}

\date{today}
\begin{abstract}
An exact solution is found for the problem of the center-of-band
($E=0$) anomaly  in the one-dimensional (1D) Anderson model of
localization.  By deriving and solving an equation for the
generating function $\Phi(u,\phi)$ we obtained an exact expression
in quadratures for statistical moments $I_{q}=\langle |\psi_{E}({\bf
r})|^{2q}\rangle$ of normalized wavefunctions $\psi_{E}({\bf r})$
which show violation of one-parameter scaling and emergence of an
additional length scale   at $E\approx 0$.

\end{abstract}
 \maketitle

--{\it Introduction.} Anderson localization (AL) enjoys an unusual
fate of being a subject of advanced research during a half of
century. The seminal paper by P.W.Anderson \cite{Anderson} opened up
a direction of research on the interplay of quantum mechanics and
disorder which is of fundamental interest up to now
\cite{Mirlin2008}. The one-dimensional (1D) tight-binding model with
diagonal disorder --the Anderson model (AM)-- which is the simplest
and the most studied model of this type, became a paradigm of AL:
\begin{equation}
\label{Ham}
\hat{H}=\sum_{i}\varepsilon_{i}\,c^{\dagger}_{i}c_{i}-\sum_{i}t_{i}\left(\,c^{\dagger}_{i}c_{i+1}+
c^{\dagger}_{i+1}c_{i}\right).
\end{equation}
In this model the hopping integral is deterministic $t_{i}=t=1$ and
the on-site energy $\varepsilon_{i}$ is a random Gaussian variable
uncorrelated at different sites and characterized by the variance
$\langle(\delta\varepsilon_{i})^{2}\rangle=w$.

The best studied is the continuous limit of this model in which the
lattice constant $a\rightarrow 0$ at $ta^{2}$ remaining finite
\cite{Ber, Mel}. There was also a great deal of activity \cite{
Pastur} aimed at a rigorous mathematical description of 1D AL.
However, despite considerable efforts invested, some subtle issues
concerning 1D AM still remain unsolved. One of them is the effects
of commensurability between the de-Broglie wavelength $\lambda_{E}$
(which depends on the energy $E$) and the lattice constant $a$.

It was known for quite a while \cite{Wegner, Derrida} that at weak
disorder $w\ll 1$ the Lyapunov exponent takes anomalous values at
the ratio $f=\frac{2a}{\lambda_{E}}$ equal to $\frac{1}{2}$ and
$\frac{1}{3}$ (compared to those at  $f$ beyond the window of the
size $w$ around $f=\frac{1}{2}$ and $f=\frac{1}{3}$). The Lyapunov
exponent sharply {\it decreases} at $f=\frac{1}{2}$ (which is
usually associated with {\it increasing} the localization length)
and may both increase or decrease at $f=\frac{1}{3}$ depending on
the third moment of the on-site energy distribution \cite{Derrida}.
 More recently \cite{Titov, AL}
it was found that the statistics of conductance in 1D AM is
anomalous at the center of the band that corresponds to
$f=\frac{1}{2}$. We want to stress that all these anomalies were
observed for the AM Eq.(\ref{Ham}) in which the on-site energy
$\varepsilon_{i}$ is random. This Hamiltonian does not possess the
{\it chiral symmetry} \cite{Dyson, Mirlin2008} which is behind the
statistical anomalies at the center of the band $E=0$ in the {\it
Lifshitz model} described by Eq.(\ref{Ham}) with the deterministic
$\varepsilon_{i}=0$ and a random hopping integral $t_{i}$. Thus the
statistical anomaly at $f=\frac{1}{2},\frac{1}{3}$  raises a
question about a {\it hidden symmetry} that does not merely reduce
to the two-sublattice division \cite{Dyson, AL, Mirlin2008}.

A similar phenomenon may occur in dynamical systems.   An elegant
analogy between the 1D localization and the classical system of
kicked oscillator was studied  in Ref.\cite{Izrail}. According to
this analogy the energy-dependent de-Broglie wavelength
$\lambda_{E}$ is encoded in the frequency of the oscillator and the
lattice constant $a$ determines the period of the $\delta$-function
"kicks" of the external force, their amplitude being proportional to
disorder.

The interest to one-dimensional Anderson localization is greatly
increased recently after several groups reported about successful
experiments on localization of cold atoms \cite{Aspekt, Ingusch},
where even tiny details of localized wavefunctions were observed.
Kicked rotors and kicked oscillator can also be realized in systems
of cold atoms \cite{Kick-rot-real}.

There are numerous questions concerning physics behind the
anomalies. One of puzzles is the sign of the variation of the
Lyapunov exponent which corresponds to {\it weaker} localization at
$f=\frac{1}{2}$. Such a tendency can be considered as a remnant of
the chiral symmetry spoiled by fluctuating on-site energy. There is,
however, a completely different view on the problem which predicts
the {\it stronger} localization at the band center. It involves the
notion of {\it Bragg mirrors} \cite{MuzKh} created by disorder
realizations with alternating on-site energies which double   the
period, at least locally. A possible resolution of this conflict
between different mechanisms of the center-of-band anomaly could be
a typical wavefunction sketched in Fig.1. It contains {\it two}
length scales: one of them $\ell_{{\rm loc}}$ which is somewhat
larger than the localization length $\ell$ away from the anomaly, is
due to remnants of the chiral symmetry, while the other, much
smaller one $d\ll \ell_{{\rm loc}}$ (but $d\gg a$), is due to the
formation of the Bragg mirror fluctuation.
\begin{figure}[h]
\includegraphics[width=8cm, height=8cm, angle=-90]{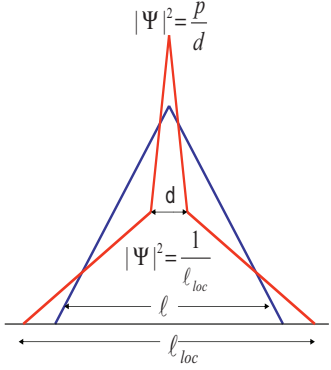}
\caption{ The cartoon of the standard (blue) and the $E=0$ typical
state (red). At the anomaly the second scale $d$ emerges.}
\end{figure}
If the weight of the narrow peak $p\ll 1$ is small, the statistical
moments $I_{q}=\langle |\psi({\bf r})|^{2q}\rangle$ of the
normalized wavefunctions $\psi({\bf r})$ with relatively small $q$
will follow the standard \cite{Mirlin2000} behavior ($L$ is the
length of the chain):
\begin{equation} \label{Ist}I_{q}^{({\rm
st})}=
%\frac{(q-1)!}{\ell_{{\rm loc}}^{q-1}\,L},
(q-1)!\,\ell_{{\rm loc}}^{1-q}\,L^{-1}.
\end{equation}
while at large $q$, the higher and more narrow peak will dominate in
$I_{q}$, leading to the $d^{-(q-1)}$ instead of $\ell_{{\rm
loc}}^{-(q-1)}$ behavior of moments. The simplest crossover between
the two regimes is described by
\begin{equation}
\label{cross} I_{q}=I_{q}^{({\rm st})}\,\left[1+p\,\left(
\frac{p\ell_{{\rm loc}}}{d}\right)^{q-1}\right].
\end{equation}
Eq.(\ref{cross}) can be obtained from the following qualitative
arguments. The average moment is the sum of two contributions. The
first one leading  to the standard moment Eq.(\ref{Ist}), is equal
to the $q$-th power of the typical amplitude $|\psi({\bf
r})|^{2}\sim 1/\ell_{{\rm loc}}$ inside the localization radius but
outside of the narrow peak,  multiplied by the probability $\sim
\ell_{{\rm loc}}/L$ that the observation point ${\bf r}$ falls
inside this region. The second contribution $(p/d)^{q}\,d/L$ arises
when with small probability $d/L$ the observation point falls inside
the narrow peak where the amplitude $|\psi({\bf r})|^{2}\sim p/d$.
It is this contribution which corresponds to the $d$-dependent term
in Eq.(\ref{cross}).

In general, the information about a typical shape of localized
wave functions is encoded in statistical moments $I_{q}$.
In this Letter we solve exactly the problem of statistical
moments $I_{q}$ at $E=0$ for the 1D Anderson disordered chain
Eq.(\ref{Ham}) of the length $L\rightarrow\infty$ and show that the
behavior Eq.(\ref{cross}) indeed emerges.

-- {\it Generating function (GF), moments of $|\psi|^{2}$ and the probability distribution function (PDF) of phase.} %The
Moments
$I_{q}$ of normalized eigenfunctions with integer $q>1$
\begin{widetext}
\begin{equation}
\label{mom} I_{q}({\bf r})=\langle|\psi_{E}({\bf
r})|^{2q}\rangle=\frac{2\pi}{L\,
(q-2)!}\int_{0}^{\pi}d\phi\,\cos^{2q}(\phi)\int_{0}^{\infty}dz\,z^{q-2}\,\Phi_{r-1}(z,\phi-k)\,
\Phi_{N-r}(z,-\phi-k),\,\,\,\,\,(E=2\cos k).
\end{equation}
\end{widetext}
can be expressed \cite{foot} in terms of a {\it generating function}
$\Phi_{j}(u,\phi)$ on the lattice site $j$. The starting point of
our analysis is the recursive equation for GF which can be derived
elementary \cite{LongKY}starting from Eq.(\ref{Ham}) as well as
using the super-symmetry method \cite{OsK}:
\begin{widetext}
\begin{equation}
\label{rec-TM} \Phi_{j+1}(z,\phi)=\frac{\sin
k\,e^{-z\,\cos^{2}\phi}}{\sqrt{2\pi\,w}\,\cos^{2}\phi}\int_{0}^{\pi}d\phi'\,{\rm
exp}\left[-\frac{\sin^{2}k}{2w}\,(\tan\phi-\tan\phi')^{2}
\right]\,\Phi_{j}\left(
z\,\frac{\cos^{2}\phi}{\cos^{2}\phi'},\;\phi'-k\right),\;\;\;\;\Phi_{j=0}(z,\phi)=\delta(\phi-\pi/2).
\end{equation}
\end{widetext}
This equation is {\it exact} and holds both for weak and strong
disorder controlled by the parameter $w$ for any energy $E=E(k)$
parametrized by $E(k)=2 \cos k$.

The way the variables $z$ and $\phi$ enter Eq.(\ref{mom})  suggests
their physical meaning\cite{Wegner}: they determine the values of a
wave function $\psi(i+1)$ and $\psi(i)$ on a link $\{i,i+1\}$:
\begin{equation}
\label{physmean}
\psi(i)=\sqrt{z_{i}}\,\cos(\phi_{i}),\;\;\;\;\psi(i+1)=\sqrt{z_{i}}\,\cos(\phi_{i}-k).
\end{equation}
It is remarkable that both the "elementary" \cite{LongKY} and the
"super-symmetric" \cite{OsK} derivations of eigenfunction statistics
involve naturally the {\it two} link variables Eq.(\ref{physmean}).
In contrast to the moments $I_{q}$, the Lyapunov exponent
\begin{equation}
\label{Lyap}
\gamma=\lim_{N\rightarrow\infty}\frac{1}{N}\sum_{i=1}^{N}\ln\left(
\frac{\psi(i+1)}{\psi(i)}\right)
\end{equation}
depends on only one of the two sets of variables, $\phi$ ($\psi(i+1)/\psi(i) = \cos(\phi_{i}-k)/\cos(\phi_{i})$), which determines completely its
statistics \cite{Wegner, Derrida}. That is why the problem of
moments is more complicated and more general than that of the
Lyapunov exponent.

The integrand in Eq.(\ref{mom}) is {\it bi-linear} in
$\Phi$. This effectively takes into account the boundary conditions
at the {\it two} ends of the chain \cite{Mirlin2000} which is
necessary to describe the {\it normalized} {\it eigen}functions. In
contrast to that in the problem of Lyapunov exponent \cite{Wegner}
one considers essentially a {\it semi-infinite} chain and does not
require of the solution to Eq.(\ref{Ham})  to be an eigenfunction.

Despite the fact that $\Phi_{j}(z,\phi)$ is {\it not} the joint PDF
of $z$ and $\phi$, its descender $\Phi_{j}(z=0,\phi)$ is the PDF of phase:
\begin{equation}
\label{P-phi} \Phi_{j}(z=0,\phi)=P_{j}(\phi),
\;\;\;\;\int_{0}^{\pi}P_{j}(\phi)\,d\phi=1.
\end{equation}
This statement can be formally proven \cite{LongKY} but the  key
properties of PDF, the  positivity of $P_{j}(\phi)$ and the
conservation of normalization, are easily seen directly from
Eq.(\ref{rec-TM}) and the boundary condition
$\Phi_{0}(\phi)=\delta(\phi-\pi/2)$.

--{\it Evolution equation for weak disorder.} Eq.(\ref{rec-TM}) is
valid for an arbitrary strength of disorder. However, the anomaly we
are going to study is sharp only at weak disorder and is rounded off
as disorder increases. For weak disorder when the localization
length $\ell=\frac{2a\,\sin^{2}k}{w}$ is large compared to the the
lattice constant $a$ one can reduce Eq.(\ref{rec-TM}) to a partial
differential equation (DE) of the Fokker-Planck type, where the
coordinate $x=j\,a/\ell$ along 1D chain plays a role of time and the
{\it two-dimensional} space of variables $u=\ell z$ and $\phi$
stands for the coordinate space.

In the first order in $a/\ell\ll 1$ one obtains by a proper
expansion in Eq.(\ref{rec-TM}):
\begin{equation}
\label{op}
\Phi_{j+1}(u,\phi)=\left(1+\frac{a}{\ell}\,\left[\hat{{\cal L
}}(\phi)-c_{1}(\phi)\,u\right]\right)\,\Phi_{j}(u,\phi-k),
\end{equation}
where the evolution operator
$\left(1+\frac{a}{\ell}\,\left[\hat{{\cal L
}}(\phi)-c_{1}(\phi)\,u\right]\right)$ contains the differential
part
\begin{eqnarray}
\label{dif} \hat{{\cal
L}}(\phi)&=&c_{2}(\phi)\,u^{2}\partial^{2}_{u}+c_{3}(\phi)\,(u\partial_{u}-1)\\
\nonumber &+&c_{4}(\phi)\, u\partial_{u}\partial_{\phi}
+c_{5}(\phi)\,\partial_{\phi}+c_{6}(\phi)\,\partial^{2}_{\phi} \,
\end{eqnarray}
and $c_{i}(\phi)$ are certain linear combinations of $1$,
$\sin(2\phi)$, $\cos(2\phi)$, and $\sin(4\phi)$, $\cos(4\phi)$.
%\begin{eqnarray}
%\label{c} c_{1}(\phi)&=&\frac{1}{2}(1+\cos(2\phi)),
%c_{2}(\phi)=1-\cos^{2}(2\phi),\nonumber \\
%c_{3}(\phi)&=&-(1-\cos(2\phi)-2\cos^{2}(2\phi)),\nonumber \\
%c_{4}(\phi)&=&\sin(2\phi)(1+\cos(2\phi)),\nonumber \\
%c_{5}(\phi)&=&-\frac{3}{2}\sin(2\phi)(1+\cos(2\phi)),\nonumber \\
%c_{6}(\phi)&=&\frac{1}{4}(1+\cos(2\phi))^{2}.
%\end{eqnarray}

The formal reason for the center-of-band anomaly at
$k=\frac{\pi}{2}$ (as well as of the weaker anomaly at any
$k=\pi\,p/q$, where $p,q$ are positive integers) is the shift by $k$
of the $\phi$ argument in r.h.s. of Eq.(\ref{op}). Because of this
shift and the periodicity $\Phi_{j}(u,\phi)=\Phi_{j}(u,\phi+\pi)$ ,
one has to apply the evolution operator $q$ times in order to get a
closed recursive equation which expresses $\Phi_{j+q}(u,\phi)$ in
terms of $\Phi_{j}(u,\phi)$ and its derivatives. For weak disorder
and not very large  $q\ll \ell/a$, one can expand
$\Phi_{j+q}-\Phi_{j}\approx
(aq/\ell)\,\partial\Phi(u,\phi;x)/\partial x$, where we introduce a
function $\Phi(u,\phi;x)=\Phi_{\ell\,x/a}(u,\phi)$ of a continuous
dimensionless coordinate $x=j\,a/\ell$. Thus in the lowest order in
$a/\ell$ we obtain for $k=\pi p/q$:
\begin{eqnarray}
\label{lin} \partial_{x} \Phi=\left[\sum_{s=0}^{q-1}\hat{{\cal
L}}(\phi-\frac{s\,\pi
p}{q})-u\sum_{s=0}^{q-1}c_{1}(\phi-\frac{s\,\pi p}{q}) \right]\Phi.
\end{eqnarray}
The sum over $s$ arises because the small corrections to the
evolution operator proportional to $a/\ell$ add up in the product of
$q$ evolution operators, each time entering with a shift
$c_{i}(\phi)\rightarrow c_{i}(\phi-k)$ according to  Eq.(\ref{op}).
The crucial point for emergence of anomaly at $k=\frac{\pi}{2}$
($q=2$, $p=1$) is the identity:
\begin{equation}
\label{id1} \sum_{s=0}^{q-1}e^{2i\phi-2i s\,\pi p/q}=0,\;\;
\sum_{s=0}^{q-1}e^{4i\phi-4i s\,\pi p/q }=\left\{\begin{matrix}0, &
q>2\cr q\,e^{4i\phi},& q=2\cr
\end{matrix} \right.
\end{equation}
One observes that at $k=\pi p/q$ with all $q$ but $q=2$ the
$\phi$-dependent terms disappear from the r.h.s. of Eq.(\ref{lin}).
At $k=\frac{\pi}{2}$, however, one obtains the anomalous,
$\phi$-dependent, evolution equation. It appears to have a nice
$SL(2)$ group structure:
\begin{eqnarray}
\label{anomal-equation} &&
\partial_{x}\Phi(u,\phi;x)=\{\hat{L}_{1}^{2}+\hat{L}_{3}^{2}-u\}\,\Phi(u,\phi;x),\\
&&\hat{L}_{1}=\cos\theta\,\partial_{\theta}+\sin\theta\,u\partial_{u},\;\;\;\hat{L}_{3}=\partial_{\theta},\;\;\;
(\theta=2\phi),\nonumber
\end{eqnarray}
where $\hat{L}_{1}$ and $\hat{L}_{3}$  and
$\hat{L}_{2}=[\hat{L}_{3},\hat{L}_{1}]=-\sin\theta\,\partial_{\theta}+\cos\theta\,u\partial_{u}$
form a closed $sl(2)$ algebra.

Note that Eq.(\ref{anomal-equation}) contains all the known
particular results. For instance, omitting all the $\phi$-dependent
terms one obtains the standard equation for GF away from the anomaly
\cite{Mirlin2000} which allows for the $\phi$- and $x$-independent
({\it zero-mode}) solution \cite{Wegner, Derrida}:
\begin{equation}
\label{stand} \Phi^{({\rm
st})}(u)=\frac{2}{\pi}\,\sqrt{u}\,K_{1}(2\sqrt{u}).
\end{equation}
Alternatively, in agreement with Eq.(\ref{P-phi}), by setting $u=0$
in Eq.(\ref{anomal-equation}) one arrives at the second order ordinary DE
for the non-trivial phase-distribution function $P_{0}(\phi)$ at the
$k=\frac{\pi}{2}$ anomaly with the zero-mode solution:
\begin{equation}
\label{P0}
P_{0}(\phi)=\Phi(0,\phi)=\frac{C}{\sqrt{3+\cos(4\phi)}},\;\;\;\; C=\frac{4\sqrt{\pi}}{\Gamma^{2}\left(\frac{1}{4}\right)} \, ,
\end{equation}
resulting in the anomaly of the Layapunov exponent \cite{Wegner, Derrida}:
\begin{equation}
\label{Lyapunov} \frac{\gamma(E=0)}{\gamma(E\neq 0)}=
\int_{0}^{\pi}(1+\cos(4\phi))\,P_{0}(\phi)=\frac{8\,\Gamma^{2}\left(
\frac{3}{4}\right)}{\Gamma^{2}\left( \frac{1}{4}\right)}\approx
0.9139.
\end{equation}
Derivation of  Eq.(\ref{anomal-equation}), and its \emph{exact}
solution is the main result of this Letter.

--{\it Separation of variables and the zero-mode solution. } The
variables $u$ and $\phi$ are entangled in
Eq.(\ref{anomal-equation}). However, there is a hidden symmetry
which allows to separate variables in this equation, provided that
the term $\partial_{x}\,\Phi(u,\phi;x)=0$. This {\it zero mode}
solution is sufficient to describe anomalous eigenfunction
statistics in a very long chain $L\gg \ell$ far from its ends.

"Correct variables" $\xi$ and $\eta$ are suggested by
Eq.(\ref{physmean}):
\begin{equation}
\label{corr-varia}
\xi=u\,\cos^{2}\phi,\;\;\;\;\;\eta=u\,\sin^{2}\phi.
\end{equation}
Defining also the "correct function":
\begin{equation}
\label{corr-fu}
\tilde{\Phi}(\xi,\eta)=\frac{(\xi\eta)^{\frac{1}{4}}}{(\xi+\eta)}\,\Phi(u(\xi,\eta),\,\phi(\xi,\eta)),
\end{equation}
one casts the zero-mode variant of Eq.(\ref{anomal-equation}) in the
form of the Schroedinger equation:
\begin{equation}
\label{Schroe} \left[\hat{H}(\xi)+\hat{H}(\eta)
\right]\,\tilde{\Phi}=0,\;\;\;\;\hat{H}(\xi)=-\partial^{2}_{\xi}-\frac{3}{16\,\xi^{2}}+\frac{1}{4\xi}.
\end{equation}
Note that the singular operator $\hat{H}(\xi)$ is not Hermitian for
generic wave function. Its spectrum  is {\it continuous} and, in
general, complex. The zero-mode solution corresponds to a zero sum
of the two eigenvalues ($\pm\Lambda$) of the 1D Hamiltonians
$\hat{H}(\xi)$ and $\hat{H}(\eta)$. Thus the solution to
Eq.(\ref{Schroe}) emerges as an integral over a continuous variable
$\lambda \propto 1/\sqrt{\Lambda}$ which can be taken real without
loss of generality \cite{LongKY}. The integrand involves the product
of two eigenfunctions $\Psi_{\Lambda}(\xi)$ and
$\Psi_{-\Lambda}(\eta)$, and an arbitrary function $C(\lambda)$.
Yet, one can find this function $C(\lambda)$ uniquely \cite{LongKY}
using the conditions of (i) smoothness of $\Phi(u,\phi)$ at $\phi=0$
and $\phi=\frac{\pi}{2}$ and (ii) normalization of the phase
distribution function $P_{0}(\phi)=\Phi(u=0,\phi)$:
\begin{widetext}
\begin{equation}
\label{solu-F} \Phi_{{\rm
an}}(u,\phi)=\frac{u^{\frac{1}{2}}}{\Gamma^{4}\left(
\frac{1}{4}\right)\,|\cos\phi\,\sin\phi|^{\frac{1}{2}}}\int_{0}^{\infty}
d\lambda\, \frac{\Gamma\left(\frac{1}{4}+\epsilon\lambda \right)
\,\Gamma\left(\frac{1}{4}+\bar{\epsilon}\lambda\right)}{\lambda^{\frac{3}{2}}}\,
{\rm Re}\,\left[W_{-\epsilon\lambda,
\frac{1}{4}}\left(\frac{\bar{\epsilon}\,\xi}{4\lambda}\right)\,W_{-\bar{\epsilon}\lambda,
\frac{1}{4}}\left(\frac{\epsilon\,\eta}{4\lambda}\right)\right],
\end{equation}
\end{widetext}
where $W_{\alpha,\frac{1}{4}}(x)$ is the Whittaker function,
%(also expressed via the parabolic cylinder function),
$\Gamma(x)$ is the Euler Gamma-function, and $\epsilon=e^{i\pi/4}$,
$\bar{\epsilon}=e^{-i\pi/4}$.

Eq.(\ref{solu-F}) is the main analytic result of the Letter.

--{\it Moments of normalized eigenfunctions.}  A convenient way to
present the results is to plot the {\it reduced moments}
$R_{q}=I_{q}(E=0)/I_{q}(E\neq 0)$,
\begin{equation}
\label{q-mom} R_{q}=
C_{q}\int_{0}^{\infty}du\int_{0}^{\pi/2}d\phi\,\cos^{2q}(\phi)\,u^{q-2}\,
\Phi_{{\rm an}}^{2}(u,\phi).
\end{equation}
Here  $I_{q}(E\neq 0)=L^{-1}\,(q-1)!\,\ell^{1-q}$ are the moments
away from the anomaly, where $\Phi(u,\phi)=\Phi^{({\rm st})}(u)$ is
given by Eq.(\ref{stand}), and
$C_{q}=\frac{\pi\,4^{q}}{(q-1)!(q-2)!}$. Using the solution
Eq.(\ref{solu-F}) we evaluated the reduced moments $R_{q}$
numerically up to $q=10$.    The results are given in Fig.2.
\begin{figure}[]
\includegraphics[width=6cm, height=8
cm,angle=-90]{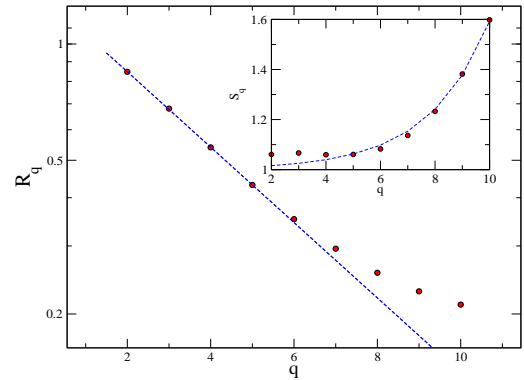} \caption{(color line)Reduced moments
$R_{q}$ (red points) in the log-linear scale. The dashed line is the
exponential fit $R_{q}=(\ell/\ell_{{\rm loc}})^{q-1}$ with $l_{{\rm
loc}}(E=0)/\ell=1.252$. In the insert: The excess factor (red
points). The dashed line is a fit by Eq.(\ref{cross}) with $p\approx
0.01$ and $l_{{\rm loc}}(E=0)/d\approx 160$.}
\end{figure}
One can see that  at $E=0$ the moments $R_{q}\approx
(\ell/\ell_{{\rm loc}})^{q-1}$ with small $q$ follow Eq.(\ref{Ist}),
albeit with a localization length $\ell_{{\rm loc}}$ larger than
that away from the   anomaly. The best exponential fit of moments
with $q<6$ gives $\ell_{{\rm loc}}/\ell\approx 1.252$. This reflects
the same tendency as Eq.(\ref{Lyapunov}). However, larger moments
are significantly greater than the prediction of one-parameter
scaling Eq.(\ref{Ist}). The {\it excess factor}
$S_{q}=I_{q}(E=0)/I_{q}^{({\rm st})}$ which should be compared with
that in the square brackets of Eq.(\ref{cross}), is plotted in the
insert of Fig.2. A comparison with Eq.(\ref{cross}) shows a very
satisfactory (for a crude qualitative interpretation in terms of two
scales sketched in Fig.1) agreement for moments up to $q=10$ which
can be interpreted as an emergence of a very narrow (but still much
wider than the lattice constant) peak in an "average" eigenfunction
at the anomaly.

In conclusion, we solved exactly the problem of statistical moments
$I_{q}$ of the amplitude $|\psi_{E}({\bf r})|^{2}$ of random wave
functions in the 1D Anderson model at energies $E\approx 0$. It is
shown that the statistics of such wavefunctions is anomalous which
anomaly does not merely reduce to the variation of the localization
length or the Lyapunov exponent. The enhancement of the localization
length $\ell_{{\rm loc}}/\ell\approx 1.252$ derived from
$I_{q}\propto \ell_{{\rm loc}}^{1-q}$ with $q<6$
 is different from that obtained from the inverse Lyapunov exponent
$\gamma(E\neq 0)/\gamma(E=0)\approx 1.094$. This fact together with
the anomalous enhancement of moments with large $q>6$ implies a
significant change of the form of the typical eigenfunction at
$E\approx 0$ which requires more than one characteristic length for
its description.

 --{\it Acknowledgement.}
 We appreciate stimulating discussions with A.Agrachev, B.L.Altshuler
Y.V.Fyodorov, A.Kamenev, A.Ossipov, O.Yevtushenko and a support from
RFBR grant 09-02-1235 (V.Y.). We are especially grateful to E.Cuevas
and D.N.Aristov for a help in numerical calculations.

Part of the work was done during visits of V.Y.
to the Abdus Salam International Center for Theoretical Physics which support is highly acknowledged.\\


\begin{thebibliography}{99}
\bibitem{Anderson} P.W.Anderson, Phys.Rev. {\bf 109}, 1492 (1958).
\bibitem{Mirlin2008} F.Evers and A.D.Mirlin, Rev.Mod.Phys. {\bf 80}, 1355 (2008).
\bibitem{Ber} V.L.Berezinskii, Zh.Exp.Teor.Fiz. {\bf 65}, 1251
(1973)[Sov.Phys.JETP {\bf 38}, 620 (1974)].
\bibitem{Mel} V.I.Melnikov, JETP Lett. {\bf 32}, 225 (1980).
\bibitem{Pastur} I.M.Lifshitz, S.A.Gredeskul, and L.A.Pastur,
{\it Introduction to the theory of disordered systems} (Wiley, N.Y., 1988).
\bibitem{Wegner}M.Kappus and F.Wegner, Z.Phys. B {\bf 45}, 15 (1981).
\bibitem{Derrida} B.Derrida and E.Gardner, J.Phys. (Paris) {\bf 45}, 1283
(1984).
\bibitem{Titov} H.Schomerus and M.Titov, Phys.Rev. B {\bf 67},
100201(R) (2003).
\bibitem{AL} L.I.Deych, et al., Phys.Rev.Lett. {\bf 91}, 096601
(2003).
\bibitem{Dyson} F.J.Dyson, Phys.Rev. {\bf 92}, 1331 (1958).
\bibitem{Izrail} L.Tessieri and F.M.Izrailev, Phys.Rev. E {\bf 62},
3090 (2000).
\bibitem{Aspekt} J.Billy, V.Josse,Z.Zuo, A.Bernard, B.Hambrecht, P.Lugan, D.Clement,
L.Sanches-Palencia, P.Bouyer, and A.Aspect, Nature (London) {\bf
453}, 891 (2008).
\bibitem{Ingusch} G.Roati, C.D'Errico, L.Fallani, M.Fattori, C.Fort, M.Zaccanti,
G.Modugno, and M.Inguscio, Nature (London) {\bf 453}, 895 (2008).
\bibitem{Kick-rot-real} F.L.Moore, J.C.Robinson, C.F.Bharucha, S.Sundaram, and M.G.Raizen,
 Phys.Rev.Lett. {\bf 75},
4598 (1995).
\bibitem{MuzKh} B.A.Muzykantskii and D.E.Khmelnitskii, Phys.Rev.B {\bf
51}, 5480 (1995); I. E. Smolyarenko and B. L. Altshuler, Phys. Rev.
B {\bf 55}, 10451 (1997); V.M.Apalkov, M.E.Raikh, and B.Shapiro,
Phys.Rev.Lett. {\bf 92}, 066601 (2004).
\bibitem{foot} The analogous expression away from anomaly is given
in Ref.\cite{Mirlin2000}, Sec.3.1.
\bibitem{Mirlin2000} A.D.Mirlin, Phys.Rep. {\bf 326}, 259 (2000).
\bibitem{LongKY} V.E.Kravtsov and V.I.Yudson, arXiv:1011.1480 (unpublished)
\bibitem{OsK} A.Ossipov and V.E.Kravtsov, Phys.Rev. B {\bf 73}, 033105 (2006).
\end{thebibliography}
\end{document}